# *Caveats* for using statistical significance tests in research assessments

*Jesper W. Schneider*

*Danish Centre for Studies in Research and Research Policy,*

*Department of Political Science & Government, Aarhus University,*

*Finlandsgade 4, DK-8200, Aarhus N, Denmark*

*jws@cfa.au.dk*

*Telephone: 0045 8716 5241*

*Fax: 0045 8716 4341*

**Abstract**
This article raises concerns about the advantages of using statistical significance tests in research assessments as has recently been suggested in the debate about proper normalization procedures for citation indicators by Opthof and Leydesdorff (2010). Statistical significance tests are highly controversial and numerous criticisms have been leveled against their use. Based on examples from articles by proponents of the use statistical significance tests in research assessments, we address some of the numerous problems with such tests. The issues specifically discussed are the ritual practice of such tests, their dichotomous application in decision making, the difference between statistical and substantive significance, the implausibility of most null hypotheses, the crucial assumption of randomness, as well as the utility of standard errors and confidence intervals for inferential purposes. We argue that applying statistical significance tests and mechanically adhering to their results are highly problematic and detrimental to critical thinking. We claim that the use of such tests do not provide any advantages in relation to deciding whether differences between citation indicators are important or not. On the contrary their use may be harmful. Like many other critics, we generally believe that statistical significance tests are over- and misused in the empirical sciences including scientometrics and we encourage a reform on these matters.

**Highlights**: We warn against the use of statistical significance tests (NHST) in research assessments. We introduce the controversial debate of NHST to the informetric community. We demonstrate some of the numerous flaws, misconceptions and misuses of NHST. We discuss potential alternatives and conclude that no "easy fixes" exist. We advocate informed judgment, free of the NHST-ritual, in decision processes.

**Acknowledgement.** The author would like to thank the three anonymous reviewers for their very useful comments on previous drafts of this article.

## 1. Introduction

In a recent article Opthof and Leydesdorff (2010; hereafter O&L) make several claims against the validity of journal and field normalization procedures applied in the so called "crown indicator" developed by the Center for Science and Technology Studies (CWTS) at Leiden University in the Netherlands. Like Lundberg (2007) before them, O&L suggest a normalization procedure based on a sum of ratios instead of a ratio of sums as used in the "crown indicator". While Lundberg (2007) and O&L give different reasons for such a normalization approach, they do commonly argue that it is a more sound statistical procedure. O&L for their part argue that, contrary to the "crown indicator", their proposed normalization procedure, which follows the arithmetic order of operations, provides a distribution with statistics that can be applied for statistical significance tests. This claim is repeated in Leydesdorff and Opthof (2010), as well as Leydesdorff et al. (2011); indeed in all these articles, Leydesdorff and co-authors distinctly indicate that significance tests are important, advantageous and somewhat necessary in order to detect "significant" differences between the units assessed.

O&L's critique and proposals are interesting and they have raised a heated but needed debate in the community (e.g., Bornmann, 2010; Moed, 2010; Spaan, 2010; Van Raan et al., 2010; Gingras & Larivière, 2011; Larivière & Gingras, 2011; Waltman et al., 2011a). We are sympathetic to the claims that sums of ratios are advantageous, nevertheless, on at least one important point we think that O&L's claims are flawed, and that is the role of statistical significance tests.

The authors seem to ignore the numerous criticisms raised against statistical significance tests throughout various decades in numerous empirical fields within the social, behavioral, medical and life sciences, for example in psychology and education (Rozeboom, 1960; Bakan, 1966; Carver, 1978; Meehl, 1978; 1990; Oakes, 1986; Cohen, 1990; 1994; Gigerenzer, 1993; Schmidt & Hunter, 1997), sociology (Morrison & Henkel, 1969), economics (McCloskey, 1985; McCloskey & Ziliak, 1996), clinical medicine and epidemiology (Rothman, 1986; Goodman, 1993; 2008; Stang, Poole & Kuss, 2010), as well as statistics proper (Berkson, 1938; Cox et al., 1977; Kruskal, 1978; Guttman, 1985; Tukey, 1991; Krantz, 1999), and recently Marsh, Jayasinghe and Bond (2011) in this journal, to name just a few non-Bayesian critical works out of literally hundreds. Unawareness of the criticisms leveled against significance tests seems to be the standard in many empirical disciplines (e.g., Huberty & Pike, 1999). For many decades the substantial criticisms have been neglected. A fact Roozeboom (1997) has called a "sociology-of-science wonderment" (p. 335). Only recently, at least in some disciplines, e.g., medicine, psychology and ecology, has the criticism slowly begun to have an effect on some researchers, journal editors, and in guidelines and textbooks, but the effect is still scanty (e.g., Wilkinson et al., 1999; Fidler & Cumming, 2007).

Statistical significance tests are highly controversial. They are surrounded by myths. They are overused and are very often misunderstood and misused (for a fine overview, see Kline, 2004). Criticisms are numerous. Some point to the inherently logical flaws in statistical significance tests (e.g., Cohen, 1994). Others claim that such tests have no scientific relevance; in fact they may be harmful (e.g., Armstrong, 2007). Others have documented a whole catalogue of misinterpretations of statistical significance tests and especially the *p* value (e.g., Oakes, 1986; Goodman, 2008). Still others have documented various different misuses, such as neglecting statistical power, indifference to randomness, adherence to a mechanical ritual, arbitrary significance levels forcing dichotomous decision making, and implausible nil null hypotheses, to name some (e.g. Gigerenzer, 1993; Shaver, 1993). Rothman (1986) exemplifies the critical perspective:

> Testing for statistical significance today continues not on its merits as a methodological tool but on the momentum of tradition. Rather than serving as a thinker's tool, it has become for some a clumsy substitute for thought, subverting what should be a contemplate exercise into an algorithm prone to error (p. 445).

Alternatives and supplements to significance tests have been suggested; among these for example effect size estimations and confidence intervals, power analyses, and study replications (e.g., Kirk, 1996). Interestingly, relatively few have defended statistical significance tests and those who have, agree with many of the criticisms leveled against such tests (e.g., Abelson, 1997; Cortina & Dunlap, 1997; Chow, 1998; Wainer, 1999). The defenders do however claim that most of these failings are due to humans and that significance tests can play a role, albeit a limited one, in research. Critics will have none of this, as history testifies that the so-called limited role is not practicable, people continue to overuse, misunderstand and misuse such tests. To critics, statistical significance tests have let the social sciences astray and scientific research can and should live without them (e.g., Carver, 1978; Armstrong, 2007).

The aim of the present article is to warn against what Gigerenzer (2004) calls "mindless statistics" and the "null ritual". We argue that applying statistical significance tests and mechanically adhering to their results in research and more specifically in research assessments, as suggested by O&L, is highly problematic and detrimental to critical (scientific) thinking. We claim that the use of such tests do not provide any advantages in relation to deciding whether differences between citation indicators are important or not. On the contrary their use may be harmful. Like many other critics, we generally believe that statistical significance tests are so problematic that reform is urgently needed (see for example, Cumming, 2012).

Centered on examples mainly from O&L (Opthof & Leydesdorff, 2010), we address some of the numerous problems of such tests. It is important to emphasize that the fallacies we discuss here are extremely common in the social sciences and not distinctive for the particular article by O&L we scrutinize. To emphasize this we provide further brief examples haphazardly retrieved from the recent scientometric literature. The reason we specifically respond to O&L's article is a grave concern that such a flawed "ritual" should be used in relation to the already sensitive issue of research assessment based on citation indicators, as well as a reaction to the argument that such tests are supposed to be advantageous in that respect. But this article is not an exhaustive review of all problems and criticisms leveled at statistical significance tests; we simple do not have the space for that and several such reviews already exist (e.g., Oakes, 1986; Nickerson, 2000; Kline, 2004). Thus we only address some of the problems and controversies due to their appearance in the article by O&L. They do not come in any natural order and are intrinsically related. We have organized the article according to the problems addressed. First we outline our understanding of how O&L approach significance tests in their article. In the second section, we proceed with some of the caveats related to the use of statistical significance tests. We outline the practice of significance tests, and we discuss some of the misconceptions and misuses, including the assumption of randomness and the utility of standard errors and confidence intervals. We conclude with a summary and some recommendations for best practice, to inspiration for authors, reviewers and editors.

## 2. The conception and application of significance tests in the article by O&L

In this section we outline how O&L employ significance tests in their article, and how they seemingly consider "significance" and present their arguments for the supposed advantages of such tests. Notice, O&L's reasoning based upon statistical significance tests have briefly been criticized in Waltman et al. (2011b). In the present article we elaborate on these matters.

O&L state that if we average over the aggregate then we can "test for the significance of the deviation of the test set from the reference set" (2010, p. 424). O&L thus claim that the normalization procedure they suggest (i.e. sum of ratios) enable statistical significance testing and that the latter can decide whether citation scores deviate "significantly" from the baseline. According to O&L this is a clear advantage, and we assume that they consider this process somewhat objective. In order to compare their normalization procedure to that of CWTS, and to demonstrate the claimed advantages of significance tests, O&L produce a number of empirical examples combined with test statistics. Differences in journal normalizations are first explored in a smaller data set. Next a "real-life" data set from the Academic Medical Centre (AMC) in the Netherlands is used to compare the relative citation scores for one scientist based on the two different normalization procedures, and subsequently to compare the effects of the different normalizations on values and ranks for 232 scholars at AMC[1]. The AMC data set is further explored in Leydesdorff and Opthof (2010a; 2010b). O&L use $p$ values in connection with Pearson and Spearman correlation coefficients, as well as the Kruskal-Wallis test used with Bonferroni corrections. In the latter case the significance level is given as 5%, whereas 1% and 5% are used with the correlation statistics. Also, citation scores based on O&L's normalization approach come with standard errors of the mean in the articles. It is a central point in O&L's argument that their normalization approach produces a statistic where uncertainty (i.e., random error) can be estimated by providing standard errors and the argument goes "[i]f the normalization is performed as proposed by us, the score is 0.91 (±0.11) and therewith *not* significantly different" (p. 426). This quote is exemplary for O&L's treatment of statistical significance tests and the apparent implicit or explicit view upon "significance". First, O&L consider the question of "significance" as a dichotomous decision, either a result is significant or not. Second, their rhetoric suggest that "significance" implies importance or rather lack of importance in this case, as the "world average" citation score of 1 is treated as a point null hypothesis, and since 1 is located within the confidence limits they cannot reject the null hypothesis, concluding that there is no "significant difference" from the "world average". Notice also that O&L use standard errors as a surrogate for tests for significance by determining whether the estimated interval subsumes the "world average" citation score or not.

---

[1] Using the Academic Medical Centre for the demonstration is interesting since CWTS has produced and delivered relative citation indicators in a previous evaluation of the centre.

We claim that the approach to statistical significance testing described above is common among social scientist. Nevertheless, it requires some critical comments because it is deeply entangled in the quagmire of problems relating to significance tests and if applied as suggested by O&L in research assessments, it may distort the decision making process and have serious consequences for those assessed. The next section addresses some of these problems and controversies.

### 3. Some caveats related to statistical significance tests

In this section we address some important problems in relation to statistical significance tests on the basis of practice, arguments and claims in O&L. First we briefly discuss what statistical significance tests are and outline their ritualistic practice. Second we define effect size and statistical power. Subsequently we address some common misinterpretations of statistical significance tests, and closely related to this, we discuss the mechanical dichotomous decision process that significance tests usually lead to. The following subsection discusses some misuses of statistical significance test, especially the implausibility of most nil null hypotheses. This leads to a discussion of one of the crucial assumptions behind such tests, randomness; and finally, we address the issue of standard errors and confidence intervals and their supposed advantages compared to $p$ values.

*3.1 The purpose and practice of statistical significance test*

The dominant approach to statistical significance testing is an unusual hybrid of two fundamentally different frequentist approaches to statistical inference, Ronald. A. Fisher's "inductive inference" and Jerzy Neyman and Egon Pearson's "inductive behavior" (Gigerenzer et al., 1989). According to Gigerenzer (2004, p.588), most hybrid significance tests are performed as a "null ritual", where:

- A statistical null hypothesis of "no difference" or "zero correlation" in the population is set up, sometimes called a nil null hypothesis. Predictions of the research hypothesis or any alternative substantive hypotheses are not specified. Notice, other hypotheses to be nullified, such a directional, non-zero or interval estimates, are possible but seldom used, hence the "null ritual".
- An arbitrary but conventional 5% significance level (or lower) is used for rejecting the null hypothesis. If the result is "significant" the research hypothesis is accepted. Results are reported as $p < .05$, $p < .01$, or $p < .001$ (whichever comes next to the obtained $p$ value). Notice, other significance levels can be used.
- This procedure is always performed.

While the "null ritual" has refined aspects, these do not change the essence of the ritual, which is identical for all statistical significance tests in the frequentist tradition. Statistical significance tests in this hybrid tradition are also popularly known as null hypothesis significance tests (NHST). NHST produces a probability value ($p$ value). The definition of the $p$ value is as follows:

- The probability of the observed data, plus more extreme data across all possible random samples, *if* the null hypothesis is true, *given* randomness[2] and a sample size of *n* (i.e., the sample size used in the particular study), and all assumptions of the test statistic are satisfied (e.g., Goodman, 2008, p. 136).

The general form can be written as $p$ (Data | $H_0$). While the mathematical definition of the *p* value is rather simple, its meaning has shown to be very difficult to interpret correctly. Carver (1978), Kline (2004) and Goodman (2008) list many misconceptions about *p* values. For example, the incorrect interpretation that if *p* = .05, the null hypothesis has only a 5% chance of being true. As the *p* value is calculated under the assumption that the null hypothesis *is* true, it cannot simultaneously be a probability that the null hypothesis is false. We are not to blame for this confusion. Fisher himself could not explain the inferential meaning of his own invention (Goodman, 2008).

The individual elements of the above statement about NHST are very important, though often neglected or ignored. First, it is important to realize that *p* values are conditional probabilities that should be interpreted from an objective frequentist philosophy of probability, i.e., a relative frequency "in-the-long-run" perspective (von Mises, 1928). Because the "long-run" relative frequency is a property of all events in the collective[3], it follows that a probability applies to a collective and not a single event (Dienes, 2008). Neither do probabilities apply to the truth of hypotheses as a hypothesis is not a collective. Consequently, a *p* value is not a probability of a single result; it is a conditional probability "in the long run". This can also be inferred from the definition above "… the observed data, plus more extreme data across all possible random samples". More extreme data actually refer to results that have not happened. Thus, if we repeat the study many times by drawing random samples from the same population(s), what would happen? In reality we sample only once and relate the *p* value to the actual result!

Second, the *p* value is a conditional probability of the *data* based on the assumption that the null hypothesis is true in the population, i.e., $p$ (Data | $H_0$), and therefore not the inverse probability $p$ ($H_0$ | Data) as often believed (Cohen, 1994). The theoretical sampling distribution against which results are compared (e.g., $t$, $F$, $\chi^2$ distributions) are generated by assuming that sampling occurs from a population(s) in which the null hypothesis is exactly true. Third, randomness is a fundamental assumption, it is the *raison d'être* of NHST. Without randomness, NHST become meaningless as we cannot address sampling error, the sole purpose of such tests (Shaver, 1993). Fourth, sample size is a crucial consideration, because the *p* value is a function of effect and sample sizes, as well as spread in data (Cohen, 1990). Fifth, the result of NHST is a probability statement, often expressed as a dichotomy in terms of whether the probability was less or more than the significance level (*α*). Notice that *p* values and *α* levels are two different theoretical entities. To Fisher *p* values are a property of the data and his notion of probability relating to the study. To Neyman-Pearson *α* is a fixed property of the test not the data and their conception of error rate in the "long-run" that strikes a balance between *α*, *β* (the probability of making a Type II error), and sample size *n* (Gigerenzer et al., 1989). The conventional 5% is due to Fisher (1925). Later, in his bitter arguments with Neyman and Pearson, he would discard the conventional level and argue for reporting exact *p* values.

---
[2] We use *randomness* to include both random sampling and random assignment.
[3] The set of events that an objective probability – understood as a relative long-run frequency – applies to. Technically, the set should be infinite, but this requirement is often relaxed in practice.

*3.2 Effect Size and statistical power*

In this context, two important concepts should briefly be clarified before we continue, effect size and statistical power. An effect size is a statistic that estimates the magnitude of the result in the population (e.g., Kirk, 1996). Measures of effect size can be classified as standardized or unstandardized. Standardized measures are scale-free because they are defined in terms of the variability in the data. Some well-known standardized measures include Cohen's *d*, *r*, $R^2$ and odds ratios (e.g., Kirk, 1996; Grissom & Kim, 2005). Unstandardized measures are expressed in the original units or in terms of percentages or proportions. Effect sizes are important for at least three reasons: 1) they provide crucial information for judging the importance of a result; 2) they are important for accumulation of evidence over time and thus for meta-analysis and theory building; and 3) prior to a study, estimates of anticipated effect sizes can be used in power analyses to project adequate sample size for detecting statistically significant results (e.g., Kirk, 1996; Kline, 2004, Ellis, 2010).

The statistical power of a significance tests is the probability of rejecting the null hypothesis when it is false (Cohen, 1988). Power is the complement of *β* (1 - *β*). A statistical power analysis involves four variables: significance level (*α*), sample size (*n*), effect size and power. For any statistical model, these relationships are such that each is a function of the other three. Statistical power is affected chiefly by the size of the effect and the size of the sample used to detect it (Cohen, 1988; 1990). Bigger effects are easier to detect than smaller effects, while large samples offer greater test sensitivity than small samples. Given *α* and the anticipated effect size, we can determine the sample size needed for detecting a statistically significant effect with a certain likelihood (i.e. power) when there is an effect there to be detected. Statistical power is particularly important when there is a true difference or association in the population. The test must be powerful enough to detect such differences or associations. Otherwise, a non-significant result would simply mean that a Type II error has been committed.

*3.3 Some common misinterpretations of statistical significance tests*

Despite frequent warnings in the literature, statistical significance is too often conflated with the practical or theoretical importance of empirical results. In a recent survey of management research, Seth et al. (2009, p. 7-8) found that 90% of the papers did not distinguished between statistical significance and practical importance. Statistical significance is often used as the sole criterion of importance leading to ritualistic dichotomous decision behavior and thereby deemphasizing interpretations of effect sizes (e.g., Scarr, 1997). A clear distinction must be made because statistically significant results are not necessarily important.

Statistical significance leads simply to a conclusion that *A* is different from *B*, or, at best, that *A* is greater than *B*, or that insufficient evidence has been found for a difference. Typically when we reject a null hypothesis of zero effect we conclude that there is a "significant" effect in the population. When we fail to reject we conclude that there is no effect. Many critics argue, however, that such mere binary decisions provide an impoverished view of what science seeks or can achieve. Kirk asks "[h]ow far would physics have progressed if their researches had focused on discovering ordinal relationships?" (1996, p. 754). What we appear to forget, however, is that statistical significance is a function of sample size and the magnitude of the actual effect (e.g., Rosenthal, 1994, p. 232). Large effect size and small sample size as well as small effect size and large sample size can both bring about statistical significance with matching $p$ values, but more disturbingly, such "significant effects" are most often treated the same way. Effect and sample sizes are rarely considered, but they should. Consider the following example from Schubert and Glänzel (1983) who suggest the $w$ statistic as a statistical significance test for differences between journal impact factors (JIF). In their example they compare two journals with impact factors 0.611 and 0.913. The $w$ statistic of 1.98 is larger than 1.96, the value corresponding to the 5% significance level, hence the authors conclude that "... the impact factors of the two journals differ significantly ..." (Schubert & Glänzel, 1983, p. 65). The important question, however, is whether this difference is important? Informed human judgment is needed for such a decision. "Human judgment" refers to the fact that decision-making in statistical inference is basically subjective, context depended and goal oriented (e.g., Bakan, 1966; Carver, 1978; Tukey, 1991). "Informed" refers to the fundamental premise of providing a sound basis upon which one can make a decision about importance. In that respect, we need to focus on effect sizes and confidence intervals, consider the research design, and perhaps most important, relate the result to former empirical findings and theoretical insights (e.g., Kirk, 1996). But when it comes to research assessments, such an intellectual base is virtually absent. It is a fundamental problem in relation to the application of statistical significance tests for comparison between citation indicators in research assessments that we basically do not know *a priori* what differences would be important. Obviously importance depends on context and goal of the assessment, as well as costs and benefits. But we do not have a substantial empirical and theoretical literature that can guide us with some anticipatory effect sizes to look for. Thus, in this example, for lack of anything better, we judge the standardized effect size in relation to Cohen's benchmarks for "small", "medium" and "large" effect sizes (Cohen, 1988). Cohen reluctantly proposed his benchmarks for statistical power analyses to help researchers guess on effect size when no other sources for estimation exist. Using his conventional definitions to interpret observed effect sizes in general is problematic and could easily lead to yet another form for "mindless" statistics. Cohen himself urged to interpret effect sizes based on the context of the data and his benchmarks should be seen as a last resort. Returning to the JIF example above, Cohen's $d$, a standardized mean difference effect size, yield an effect size around .24. According to Cohen (1988), "small" effect sizes begin around .20 for mean differences. Effect sizes lower than this are considered trivial and the benchmark for "medium-sized" effects is set to .50. Is the apparently "small" but statistically significant effect between the two journal impact factors important? The confidence interval for the effect size is -.01 to .48, which is from zero effect to almost a medium effect, can we base our decision on this level of uncertainty?

Important effect sizes, those we determine would make a difference, and accepted levels of uncertainty, should be defined before the study commences. But beware, big effects are not necessarily important effects, neither are small effects necessarily unimportant. Now the question of course is "how big is big?" Obviously the question is relative to the actual study and certainly not easy to answer. A relative citation impact of 1.05 can be "statistically significant" above the world average of 1, but we would probably not consider this result important or rather it depends on the context.

*3.4 Misuse of the term "significance" and the practice of dichotomous decisions*
One of the reasons for the widespread use of the null ritual may well be the false belief that statistical significance tests can decide for us whether results are important or not. By relying on the ritual we are conveniently relived of further pains of hard thinking about differences that make a difference (Gigerenzer, 2004). An overwhelming number of tests are produced in this mechanical fashion. But the truth is that most of them do not scrutinize the statistically significant differences found and it is likely that most differences are trivial despite the implied rhetoric (e.g., Webster & Starbuck, 1988). The rhetorical practice is often to drop the qualifier "statistical" and speak instead of "significant differences". Using the term "significance" without the qualifier certainly gives an impression of importance, but "significance" in its statistical sense means something quite different. It has a very limited interpretation specifically related to sampling error. Reporting that a result is "highly significant" simply means a "long-run" interpretation of how strong the data, or more extreme data, contradict the null hypothesis that the effect is zero in the population, given repeated random sampling with the same sample size. Whether the result is "highly important" is another question still not answered.

Nowhere in their articles do O&L use the qualifier "statistical". They continuously speak of "significance", "significantly different" or "*not* significantly different". For example, they emphasize that such a procedure (sum of ratios) "…allows us to test for significance of the deviation of the test set from the reference set" (2010, p. 424), or "… the researcher under study would show as performing significantly below the world average in his reference group, both with (0.71) or without self-citations (0.58)" (2010, p. 426). To us at least, it seems evident that "significance" to O&L somehow is conceived of as a criterion of importance and used as a dichotomous decision making tool in relation to research assessment, i.e., either the results are "significant" or not. There are countless examples in the scientometric literature of similar practice, where statistical significance is treated as *the* binary criterion for importance of results. For example in regression analysis, the importance of predictor variables or the fit of the model usually comes down to whether *t* or *F* statistics are "significant" or not at the conventional alpha levels (e.g., Stremersch, Verniers & Verhoef, 2007; Haslam et al., 2008 and Mingers & Xu, 2010 to name just a few studies that try to identify variables that predict citation impact).

Now it may be that O&L do in fact mean "statistical significance" in its limited frequentist sense relating to sampling error, which this quote could indicate "[t]he significance of differences depends on the shape of the underlying distributions and the size of the samples" (2010, p. 428); but if so, they explicitly fail to attend to it in a proper and unambiguous manner. And this merely raises new concerns, such as the plausibility of the null hypothesis, the assumption of randomness and the actual statistical power involved. We will address these questions in the following subsections.

Dichotomous decisions based on arbitrary significance levels are uninformed. Consider Table 1 in Leydesdorff & Opthof (2010a, p. 645), were the Spearman rank correlation between field normalized sum of ratios versus ratio of sums is not significant in this particular study (we assume at the 5% level). A calculation of the exact $p$ value, with $n = 7$, gives a probability of .052. To quote Rosnow and Rosentahl, "surely, God loves the .06 nearly as much as the .05" (1989, p. 1277). A rhetorical variant of this example is found in the following quote from Jacob, Lehrl and Henkel (2007, p. 125) "... citation rates in co-authorships almost reach significance ($p = 0.059$) pointing to a trend to support this assumption". Surely, the practical difference between .049 and .059 is miniscule but the quote also reveals a very common and serious misunderstanding, the so-called "inverse probability fallacy" (e.g., Carver, 1978). The $p$ value provides no direct information about the truth or falsity of the null hypothesis, conditional or otherwise. To recapture, NHST provides the $p(\text{Data} | H_0)$ and not $p(H_0 | \text{Data})$. The latter however is usually what researchers want to know. According to Cohen "[NHST] does not tell us what we want to know, and we so much want to know what we want to know that, out of desperation, we nevertheless believe that it does! What we want to know is 'Given these data, what is the probability that $H_0$ is true?' But as most of us know, what it tells us is 'Given that $H_0$ is true, what is the probability of these (or more extreme) data?" (p.997). We think it is unsophisticated to treat the "truth" as a clear-cut binary variable and ironic that a decision that makes no allowance for uncertainty occurs in a domain that purports to describe degrees of uncertainty. Remember also that frequentists are concerned with collectives not the truth of a single event. The ritualistic use of the arbitrary 5% or 1% levels induces researchers to neglect critical examination of the relevance and importance of the findings. Researchers must always report not merely statistical significance but also the actual statistics and reflect upon the practical or theoretical importance of the results. This is also true for citation indicators and differences in performance rankings. To become more quantitative, precise, and theoretically rich, we need to move beyond dichotomous decision making.

*3.5 The misuse of nil null hypotheses and the neglect of Type II errors*

Significance tests are computed based on the assumption that the null hypothesis is true in the population. This is hardly ever the fact in the social sciences. Nil null hypotheses are almost always implausible, at least in observational studies (e.g., Berkson, 1938; Lykken, 1968; Meehl, 1978; Cohen, 1990, Anderson et al., 2000). A nil null hypothesis is one which posits, in an absolute sense, no difference or no association in a parameter, and it is almost universally applied (Cohen, 1994). There will always be uncontrolled spurious factors in observational studies and it is even questionable whether randomization can be expected to exactly balance out the effects of all extraneous factors in experiments (Meehl, 1978). As a result, the observed correlation between any two variables or the difference between any two means will seldom be exactly 0.0000 to the $n$th decimal. A null hypothesis of no difference is therefore most probably implausible, and if so, disproving it is both unimpressive and uninformative (Lykken, 1968; Cohen, 1994). Add to this the sample size sensitivity of NHST (e.g., Cohen, 1990; Mayo, 2006). For example, an observed effect of $r = .25$ is statistically significant if $n = 63$ but not if $n = 61$ in a two-tailed test. A large enough sample can reject any nil null hypotheses. This property of NHST follows directly from the fact that a nil null hypothesis defines an infinitesimal point on a continuum. As the sample size increases, the confidence interval shrinks and become less and less likely to include the point corresponding to the null hypothesis. Given a large enough sample size, many relationships can emerge as being statistically significant because "everything correlates to some extent with everything else" (Meehl, 1990, p. 204)." These correlations exist for a combination of interesting and trivial reasons. Meehl (1990) referred to the tendency to reject null hypotheses when the true relationships are trivial as the "crud factor". And Tukey (1991) piercingly wrote that "… it is foolish to ask 'are the effects of *A* and *B* different?' They are always different - for some decimal place" (p. 100). What we want to know is the size of the difference between *A* and *B* and the error associated with our estimate. Consequently, a difference of trivial effect size or even a totally spurious one will eventually be statistically significant in an overpowered study. Similarly, important differences can fail to reach statistical significance in poorly designed, underpowered studies. Notice, that it is a fallacy to treat a statistically non-significant result as having no difference or no effect. For example, O&L (2010, p. 426) state that "[i]f the normalization is performed as proposed by us, the score is 0.91 (±0.11) and therewith not significantly different from the world average". Without considering statistical power and effect sizes, statistically non-significant results are virtually uninterpretable.

Consider another example. In a study on determinants of faculty research productivity, Long et al. (2009, p. 245) conclude that they cannot support their research hypothesis that doctoral students in information systems with high-status academic origins exhibit greater research productivity in terms of both quantity and quality than doctoral graduates with moderate- or low status academic origins. An *F* test indicated that differences in mean citation counts across academic origins (i.e. 108.82, 95.26 and 37.08 respectively) were not statistically significant ($p = .09$). Likewise, no "significant pairwise differences" were found. But $p = .09$ does not mean that the assumption of equality between mean citation counts exist. It does mean that the data were not inconsistent with the assumed nil null statistical hypothesis at the conventional 5% alpha level, given the actual sample size. Again we see the misconception $p(H_0 \mid Data)$. Though often emphasized that failing to reject the null hypothesis does not mean that the null hypothesis is true, when it comes to a decision this is a distinction without a difference. The practical consequence is that we act as if there was no difference in citation counts between high-status, moderate-status and low-status graduates. We suspect that similar differences in mean citation counts at $p = .05$ would have lead the authors to a supportive conclusion. But perhaps the nil null hypothesis was implausible to begin with? As in the previous example, considerations of power and effect sizes are also needed in this case in order to say anything concerning the research hypothesis.

It is important to note that Type I errors can only occur when the null hypothesis is actually true. The *p* value only exists assuming the null hypothesis to be true. Accordingly, with implausible null hypotheses, the effective rate of Type I errors in many studies may essentially be zero and the only kind of decision errors are Type II. If the nil null hypothesis is unlikely to be true, testing it is unlikely to advance our knowledge. It is more realistic to assume non-zero population associations or differences, but we seldom do that in our statistical hypotheses. In fact we seldom reflect upon the plausibility of our null hypotheses, or for that matter adequately address other underlying assumptions associated with NHST (e.g., Keselman et al., 1998). O&L do not reflect upon the plausibility of their various unstated nil null hypotheses. As far as we understand O&L, they apply Kruskal-Wallis tests in order to decide whether the citation scores of their AMC researchers are "significantly different" from unity (the world average of 1). Is the null hypothesis of exactly no difference to the $n$th decimal in the population plausible and in a technical sense true? We question that. Surely citation scores deviate from 1 at some level of precision (e.g., Berkson, 1942). Sample sizes in O&L are generally small. In the particular case of AMC researchers #117 and #118 (p. 427), their citation scores of 1.50 and .93 turns out to be *not* "significantly different" from the world average of 1. But the results are a consequence of low statistical power, i.e., small sample sizes combined with Bonferroni procedures, and we are more likely dealing with a Type II error, a failure to reject a false null hypothesis of no absolute difference. If sample sizes could be enlarged randomly, then the statistical power of the studies would be strengthened. However, if the null hypothesis is false anyway, then it is just a question of finding a sufficiently large sample to reject the null hypothesis. Sooner or later the citation scores of AMC researchers #117 and #118 will be "significantly different" from the world average. The question of course is whether such a deviation is trivial or important? Informed human judgment is needed for such decisions.

Whether the actual AMC samples are probability samples and whether they could be enlarged randomly is a delicate question which we return to in section 3.7. But consider this, if the samples are not probability samples addressing sampling error becomes meaningless, and the samples should be considered as convenience samples or apparent populations. In both cases, NHST would be irrelevant and the citation scores as they are would do, for example 1.50 and .93. Are these deviations from 1 trivial or important? Again, we are left with informed human judgment for such decisions.

*3.6 Overpowered studies*

Larivière and Gingras (2011) is generally supportive of O&L, but contrary to O&L, their analyses of the differences between the two normalization approaches are for most of them overpowered with very low *p* values. Wilcoxon-signed rank tests are used to decide whether the distributions are "statistically different" (p. 395). Like O&L, Larivière and Gingras (2011) do not reflect upon the assumptions associated with NHST, such as randomness or the plausibility of their null hypotheses. It is indeed questionable whether these crucial assumptions are met. A null hypothesis of a common median equal to zero is questionable and a (plausible) stochastic data generation mechanism is not presented. Not surprisingly, given the sample sizes involved, the differences between the two distributions are "statistically different" or "significantly different" in the words of Larivière & Gingras (2011, p. 395). The more interesting question is to what extent the results are important or, conversely, an example of the "crud factor". Wilcoxon-signed rank tests alone cannot tell us whether the differences between the two distributions are noteworthy, especially not in high-powered studies with implausible nil null hypothesis. More information is needed. Larivière and Gingras (2011) do in fact address the importance for some of their results with information extrinsic from the Wilcoxon-signed rank tests. Scores for the sum of ratios, for example, seem to be generally higher than those of the ratio of sums, and the authors reflect upon some of the potential consequences of these findings (p. 395). This is commendable. Again effect sizes and informed human judgment are needed for such inferential purposes, and it seems that Larivière and Gingras (2011) indeed use differences in descriptive statistics to address the importance of the results. Why then use Wilcoxon-signed rank tests? As a mechanical ritual that can decide upon importance? As argued this is untenable. Or as an inferential tool? In that case we should focus upon the assumption of randomness and whether this is satisfied. This is questionable. As always, sufficient power guarantees falsification of implausible null hypothesis, and this seems to be the case in Larivière and Gingras (2011). Interestingly, prior to Larivière and Gingras (2011), Waltman et al. (2011c), obtained similar empirical results comparing the two normalization approaches, albeit without involving the null ritual.

*3.7 The assumption of randomness and its potential misuse*

Statistical significance tests concern sampling error and we sample in order to make statistical inferences, either descriptive inferences from sample to population or causal claims (Greenland, 1990). Statistical inference relies on probability theory. In order for probability theory and statistical tests to work *randomness* is required (e.g., Cox, 2006). This is a mathematical necessity as standard errors and $p$ values are estimated in distributions that assume random sampling from well-defined populations (Berk & Freedman, 2003). Information on how data is generated becomes critical when we go beyond description. In other words, when we make statistical inferences we assume that data are generated by a stochastic mechanism and/or that data are assigned to treatments randomly. The empirical world has a structure that typically negates the possibility of random selection unless random sampling is imposed. Ideally, random sampling ensures that sample units are selected independently and with a known nonzero chance of being selected (Shaver, 1993). As a consequence, random samples should come from well-defined finite populations, not "imaginary" or "super-populations" (Berk & Freedman, 2003). With random sampling an important empirical matter is resolved. Without random sampling, we must legitimate that the nature or the social world produced the equivalent of a random sample or constructed the data in a manner that can be accurately represented by a convenient and well-understood model. Redner (2005), for example, suggest that citation data have a stochastic nature generated by a linear preferential attachment mechanism. Perhaps, but we are skeptical about treating social processes as genuine stochastic processes that generates the equivalent of random samples like, for example, a model of radioactive decay does in the physical world (Berk, Western & Weiss, 1995a). The social world is the domain of man-made laws, social regulations, customs, the particulars of a specific culture and the spontaneous actions of people (Winkler, 2009, p. 190-104). Also, there seems to be some debate in our community as to what theoretical distribution that best approximates a citation distribution (e.g., Viera & Gomes, 2010). We believe that randomness is best obtained through an appropriate probability sample with a well-defined population. Alternatively, data may constitute a convenience sample or an apparent population (or a census from a population) (Berk, Western & Weiss, 1995a).

*3.7.1 Convenience samples, apparent populations and "super-populations"*
Very few observational studies using inferential statistics in the social sciences clarify how data are generated, what chance mechanism is assumed, if any, or define the population to which results are generalized, whether explicitly or implicitly. Presumably, most observational studies, also in our field, are based on convenience and not probability samples (Kline, 2004). Albeit many social scientists do it, it is nevertheless a category mistake to make statistical inferences based upon samples of convenience. With convenience samples, bias is to be expected and independence becomes problematic (Copas & Li, 1997). When independence is lacking conventional estimation procedures will likely provide incorrect standard errors and *p* values can be grossly misleading. Berk and Freedman (2003) suggest that standard errors and *p* values will be too small, and that many research results are held to be statistically significant when they are the mere product of chance variation. Indeed, there really is no point in addressing sampling error when there is no random mechanism to ensure that the probability and mathematical theory behind the calibration is working consistently.

Turning to O&L, it is not at all clear in what way they assume that their samples are a product of a chance mechanism and from what well-defined populations they may have been drawn? For example, one of the cases studied in O&L concern one principal investigator (PI) from AMC. Sampling units are 65 publications affiliated with the PI for the period 1997-2006. The questions are: (a) in what sense does this data set comprise a probability sample?; and (b) how is the population defined? We assume that O&L have tried to identify all eligible publications in the database for the PI in question for the given period. Most likely, data constitutes all the available observations from the "apparent" population of publications affiliated with the PI. If so, frequentist inference based on a long-run interpretation of some repeatable data mechanism is not appropriate. There is no uncertainty due to variation in repeated sampling from the population. A counter argument could be that "the data are just one of many possible data sets that could have been generated if the PI's career were to be replayed many times over". But this does not clarify what sampling mechanism selected the career we happened to observe. No one knows, or can know. It is simply not relevant for the problem at hand to think of observations as draws from a random process when further realizations are impossible in practice and lack meaning even as abstract propositions. Adhering to a frequentist conception of probability in the face of non-repeatable data and in a non-stochastic setting seems dubious.

Neither can the set of publications identified by O&L in the specific citation database be considered a random draw from the finite population of all papers affiliated with the PI, including those external to the database. It is unlikely that the data generation mechanism can be stochastic when governed by indexing policies in one database. Most likely, the data set constitutes a convenience sample of specific publication types coincidentally indexed in the specific database. Convenience samples are often treated as if they were a random realization from some large, poorly-defined population. This unsupported assumption is sometimes called the "super-population model" (Cochran, 1953). While some authors argue that "super-populations" are justifiable for statistical significance test (e.g., Bollen, 1995), we do not find such arguments convincing for frequentist statistics with non-experimental data (see for example, Western and Jackman (1994) and Berk, Western and Weiss (1995a; 1995b) for similar views). "Super-populations" are defined in a circular way as the population from which the data would have come if the data were a random sample (Berk & Freedman, 2003). "Super-populations" are imaginary with no empirical existence, as a consequence, they do not generate real statistics and inferences to them do not directly answer any empirical questions. What draw from an "imaginary super-population" does the real-world sample we have in hand represent? We simply cannot know. Inferences to imaginary populations are also imaginary (Berk & Freedman, 2003)[4].

One could of course treat data as an apparent population. In this non-stochastic setting statistical inference is unnecessary because all the available information is collected. Nonetheless, we often still produce standard errors and significance tests for such settings, but their contextual meaning is obscure. There is no sampling error, means and variances are population parameters. Notice, population parameters can still be inaccurate due to measurement error, an issue seldom discussed in relation to citation indicators. Leaving measurement error aside for a moment, what we are left with is the citation indicator, the actual parameter, what used to be the estimated statistic. In the AMC case the indicator for the PI is .91which is below the world average of one. Is it an important deviation from the world average - maybe not?

It is somehow absurd to address standard errors, *p* values and confidence intervals with strict adherence as if these numbers and intervals were precise, when they are not. Consider the bibliometric data used for indicator calculations. They are selective, packed with potential errors that influence amongst other things the matching process of citing-cited documents (Moed, 2002). Obviously, the best possible data should be used, but this actually means that a high workload should be invested to improve bibliometric data quality. It is more than likely that the measurements that go into indicators are biased or at least not "precise" (see Adler, Ewing and Taylor (2009) for a critical review of citation data and the statistics derived from them). Notwithstanding the basic violation of assumptions, we think it is questionable to put so must trust in significance tests with sharp margins of failure when our data and measurements most likely at best are imprecise. In practice sampling is complicated and because even well-designed probability samples are usually implemented imperfectly, the usefulness of statistical inference will usually be a matter of degree. Nevertheless, this is rarely reflected upon. In practice sampling assumptions are most often left unconsidered. We believe that the reason why the assumption of randomness is often ignored is the widespread and indiscriminate misuse of statistical significance tests which may have created a cognitive illusion where assumptions behind such tests have "elapsed" from our minds and their results are thought to be something they are not, namely decision statements about the importance of the findings. One can always make inferences but statistical inferences come with restrictive assumptions and frequentist inference is not applicable in non-stochastic settings.

---

[4] Notice, there is an important difference between imaginary populations that plausibly could exist and those that could not. An imaginary population is produced by some real and well-defined stochastic process. The conditioning circumstances and stochastic processes are clearly articulated often in mathematical terms. In the natural sciences such imaginary populations are common. This is not the case in the social sciences, yet super-populations are very often assumed, but seldom justified.

*3.8 Standard errors and confidence intervals permit, but do not guarantee, better inference*

It is generally accepted among critics of statistical significance tests that interval estimates such as standard errors (SE) and confidence intervals (CI) are superior to *p* values and should replace them as a means of describing variability in estimators. If used properly, SE and CI are certainly more informative than *p* values and should replace them. They focus on uncertainty and interval estimation and simultaneously provide an idea of the likely direction and magnitude of the underlying difference and the random variability of the point estimate (e.g., Cumming, Fidler & Vaux, 2007; Coulson et al., 2010). But as inferential tools, they are bound to the same frequentist theory of probability, meaning "in-the-long-run" interpretations, and riddled with assumptions such as randomness and a "correct" statistical model used to construct the limits. A CI derived from a valid test will, over unlimited repetitions of the study, contain the true parameter with a frequency no less than its confidence level. This definition specifies the coverage property of the method used to generate the interval, not the probability that the true parameter value lies within the interval, it either does or does not. Thus, frequentist inference makes only *pre*-sample probability assertions. A 95% CI contains the true parameter value with probability .95 only before one has seen the data. After the data has been seen, the probability is zero or one. Yet CIs are universally interpreted in practice as guides to post-sample uncertainty. As Abelson puts it: "[u]nder the Law of the Diffusion of Idiocy, every foolish application of significance testing is sooner or later going to be translated into a corresponding foolish practice for confidence limits" (1997, p. 130). When a SE or CI is only used to check whether the interval subsumes the point null hypothesis, the procedure is no different from checking whether a test statistic is statistically significant or not. It is a covertly NHST.

O&L provide SE for the citation indicators in their study. These roughly correspond to 68% CI[5]. Thus, in a perfect world, the interpretation should be that under repeated realizations, the interval would cover the true citation score 68% of the time. But we have no way of knowing whether the current interval is one of the fortunate 68%, and we probably have no possibility for further replications. In addition, as pointed out in the previous section, it is indeed questionable whether the samples in O&L are genuine probability samples. If they are not, interpretation of SE becomes confused. According to Freedman (2003), "an SE for a convenience sample is best viewed as a *de minimis* error estimate: if this were—contrary to fact—a simple random sample, the uncertainty due to randomness would be something like the SE".

In the case of O&L, what seems at first to be a more informative analysis with interval estimates turns out to be a genuine significance test with dichotomous decision behavior and questionable fulfillment of crucial assumptions. SEs are used as a surrogate for NHST: "[i]f the normalization is performed as proposed by us, the score is 0.91 (±0.11) and therewith not significantly different from the world average" (p. 426). We can see from this quote, and others, that their main interest is to check whether the interval subsumes the world average of 1. In this case, it does, and consequently the implicit nil null hypothesis is not rejected. This is the null ritual and it is highly problematic.

---

[5] A CI of 95% is roughly 2 SE. SE bars around point estimates that represent means in graphs are typically one SE wide, which corresponds roughly to a 32% significance level and a 68% CI.

## 4. Summary and recommendations

Opthof and Leydesdorf (2010) provide at least one sound reason for altering the normalization procedures in relation to citation indicators, however, it is certainly not statistical significance testing. As we have discussed in the present article, statistical significance tests are highly problematic. They are logically flawed, misunderstood, ritualistically misused, and foremost mechanically overused. They only address sampling error, not necessarily the most important issue. They should be interpreted from a frequentist theory of probability and their use is conditional on restrictive assumptions, most pertinent, that the null hypothesis must be true and data generation is the result of a plausible stochastic process. These assumptions are seldom met rendering such tests virtually meaningless. The problems sketched here are well known. Criticisms and disenchantments are mounting, but changing mentality and practice in the social sciences is a slow affair, given the evidence, indeed a "sociology-of-science wonderment" as Roozeboom (1997, p. 335) phrased it.

The use of significance tests by O&L is probably within "normal science" and their practice is not necessarily more inferior to those of most others, the other examples testify to that. What has caused our response in this article is a grave ethical concern about the ritualistic use of statistical significance testing in connection with research assessments. Assessments and their byproducts, funding, promotion, hiring or sacking, should not be based on a mechanical tool known to be deeply controversial. Whether differences in rankings or impact factors between units, are important should be based on human judgment informed by numbers not by mechanical decisions based on tests that are logically flawed and very seldom based on the assumptions they are supposed to. Indeed, in their argument for changing the normalization procedures, O&L point to, what they see as flawed assessments and the real consequences they have had for individual researchers at AMC. This is laudable, but then arguing that statistical significance tests is an advancement for such assessments are problematic in our view. As we have argued, it hardly brings more objectivity or fairness to research assessments, on the contrary.

Suggested alternatives to significance tests include the use of CIs and SEs. They do provide more information and are superior to statistical significance tests and should as such be preferred. But neither CIs or SEs are a panacea for the problems outlined in this article. They are based on the same frequentist foundation as NHST

Resampling techniques (e.g., bootstrap, jackknife and randomization) are considered by some to be a suitable alternative to statistical significance tests (Diacronis & Efron, 1983). Resampling techniques are basically internal replications that recombine the observations in a data set in different ways to estimate precision, often with fewer assumptions about underlying population distributions compared to traditional methods (Lunneborg, 2000). Resampling techniques are versatile and certainly have merits. The bootstrap technique seems especially well suited for interval estimation *if* we are unwilling or unable to make a lot of assumptions about population distributions. A potential application in this area is the estimation of CIs for effect sizes and for sensitivity analyses (e.g., Colliander & Ahlgren, 2011). In the 2011 Leiden Ranking by CWTS, such interval estimations seem to have been implemented in the form of "stability intervals" for the various indicators. Notice, they are used for uncertainty estimation, not statistical inference. Indeed, the statistical inferential capabilities of resampling techniques are highly questionable. Basically, one is simulating the frequentist "in-the-long-run" interpretation but using only the data set on hand as if it were the population. If this data set is small, unrepresentative, biased, non-random or the observations are not independent, resampling from it will not somehow fix these problems. In fact resampling can magnify the effects of unusual features in a data set. Consequently, resampling does not entirely free us from having to make assumptions about the population distribution and it is not a substitute for external replication, which is always more preferable.

Other statistical tools that could inform a decision-making process when it comes to comparison and importance of results are, for example, exploratory data analyses, such as box-whiskers plot. But there are more satisfactory inferential alternatives, which contrary to NHST, do assess the degree of support that data provide for hypotheses, e.g., Bayesian inference (e.g., Gelman et al., 2004), model-based inference based on information theory (e.g., Anderson, 2008) and likelihood inference (e.g., Royall, 1997).

*4.1. Some recommendations for best practice*

Some researchers have called for a ban on NHST (e.g., Hunter, 1997). Censoring is not the way forward, but neither is status quo. What we need is statistical reforms as suggested for example by Wilkinson et al. (1999), Kline (2004), Cumming (2012). Here emphasis is on parameter estimation, i.e. effect size estimation with confidence intervals. Important publication guidelines such as APA (2010) still sanction the use of NHST, albeit with strong recommendations to report measures of effect size and confidence intervals around them (e.g. APA, 2010, p. 34).

Based on the aforementioned sources on statistical reform, here are some recommendations on data analysis practices from the frequentist perspective: 1) statistical inference only makes sense when data come from a probability sample or have been randomly assigned to treatment and control groups; 2) whenever possible take an estimation framework, starting with the formulation of research aims such as "how much?" or "to what extent?"; 3) interpretation of research results should be based on point and interval estimates; 4) calculate effect size estimates and confidence intervals to answer those questions, then interpret results based on informed judgment; 5) if statistical significance tests are used, (a) information on power must be reported, and (b) the null hypothesis should be plausible; 6) effect sizes and confidence intervals must be reported whenever possible for all effects studied, whether large or small, statistically significant or not; 7) exact *p* values should be reported; 8) it is unacceptable to describe results solely in terms of statistical significance; 9) use the word "significant" without the qualifier "statistically" only to describe results that are truly noteworthy; 10) it is the researcher's responsibility to explain why the results have substantive significance; statistical tests are inadequate for this purpose; 11) replication is the best way to deal with sampling error.

Finally, it is important to emphasize what significance tests, or CIs and SEs used for the same purpose, are *not*, and what they *cannot* do for us. They do not make a decision for us. Standard limits for retaining or rejecting our null hypothesis have no mathematical or empirical relevance, they are arbitrary thresholds. There can and should be no universal standard. Each case must be judged on its merits. Significance tests are based on unrealistic assumptions giving them limited applicability in practice. They relate only to the assessment of the role of chance and they are not very informative at that if at all. They tell us nothing about the impact of errors, and do not help decide whether any plausible substantive result is true. First and foremost, there are no magical solutions besides informed human judgment. Like the current debate on field normalization, it is time to start a debate concerning the (mis)use of statistical significance testing within our field. We encourage quantitative and statistical *thinking*, not mindless statistics. We do not think that the null ritual has much if anything that speaks for it.